\documentclass[english,aps,prepri]{revtex4}
\usepackage[T1]{fontenc}
\usepackage[latin9]{inputenc}
\usepackage{color}
\usepackage{units}
\usepackage{textcomp}
\usepackage{amsmath}
\usepackage{amssymb}
\usepackage{graphicx}
\usepackage{esint}

\makeatletter
\@ifundefined{textcolor}{}
{%
 \definecolor{BLACK}{gray}{0}
 \definecolor{WHITE}{gray}{1}
 \definecolor{RED}{rgb}{1,0,0}
 \definecolor{GREEN}{rgb}{0,1,0}
 \definecolor{BLUE}{rgb}{0,0,1}
 \definecolor{CYAN}{cmyk}{1,0,0,0}
 \definecolor{MAGENTA}{cmyk}{0,1,0,0}
 \definecolor{YELLOW}{cmyk}{0,0,1,0}
}

\usepackage{babel}
\renewcommand{\vec}[1]{\boldsymbol{#1}}

\makeatother

\usepackage{babel}
\begin{document}

\title{Topological analysis of paraxially scattered electron vortex beams}

\author{Axel Lubk12, Laura Clark2, Giulio Guzzinati2, Jo Verbeeck2}

\address{1Triebenberg Laboratory, Institute of Structure Physics, Technische
Universität Dresden, 01062 Dresden, Germany.}

\address{2 EMAT, University of Antwerp, Groenenborgerlaan 171, 2020 Antwerp,
Belgium}
\begin{abstract}
We investigate topological aspects of sub-nm electron vortex beams
upon elastic propagation through atomic scattering potentials. Two
main aspects can be distinguished: (i) Significantly reduced delocalization
compared to a similar non-vortex beam if the beam centers on an atomic
column and (ii) site symmetry dependent splitting of higher-order
vortex beams. Furthermore, the results provide insight into the complex
vortex line fabric within the elastically scattered wave containing
characteristic vortex loops predominantly attached to atomic columns
and characteristic twists of vortex lines around atomic columns. 
\end{abstract}

\keywords{Quantum optics, nonlinear optics, classical optics}

\pacs{41.85.-p 42.50.Tx 41.20.Jb 41.75.Fr}

\maketitle

\section{Introduction}

Electron vortex beams carrying orbital angular momentum (OAM) recently
gained considerable interest after the experimental demonstration
of so-called vortex electron beams in a transmission electron microscope
(TEM)\citep{Verbeeck(2010),Uchida(2010)}. Their properties were predicted
earlier as an extension to optical vortices including their charged
particle character\citep{Bliokh(2007)}. Similar to optical vortices\citep{Molina-Terriza(2007),Franke-Arnold(2008)}
several applications have been demonstrated but especially the ability
to focus electron vortex beams to the atomic scale\citep{Verbeeck(2012),Verbeeck(2011)}
and their strong interaction with matter\citep{Lloyd(2012),Schattschneider(2012)}
holds promise for atomic resolution mapping of magnetic states in
materials by means of Electron Magnetic Circular Dichroism (EMCD).

The linear wave equations governing the motion of the electron beam
allow eigensolutions to arbitrary OAM\citep{Schattschneider(2011),Bliokh(2012)},
which are experimentally created by dedicated apertures in TEM\citep{Verbeeck(2010),Verbeeck(2012),Uchida(2010)}.
Such beams must contain a phase singularity, which posses a more basic
structure only loosely connected to OAM\cite{Berry(2009)}: Topological
concepts can be used to precisely define that singularity in terms
of topological charges thereby characterizing the electron wave\citep{Berry(2007)}.
However, for the particular case of highly-localized electron vortex
beams scattering at atomic potentials almost nothing is known about
their vortex behaviour upon propagation. We will apply both (scalar)
electron wave dynamics and topological concepts in order establish
conservation laws and to study the interaction between vortex lines
and electrostatic potentials. This topological information is usually
too coarse for quantitative computation of e.g. scattering cross sections
of core-losses in Energy Electron Loss Spectroscopy (EELS). However,
it provides qualitative understanding and concepts sometimes without
solving the scattering problem itself. E.g. we will show subsequently
that wave topology is sufficient to characterize some aspects of the
wave function upon propagation, like the creation and destruction
of electron vortex beams, the splitting of higher order vortex beams,
etc. We also stress the generality and beauty of topological concepts
by appealing to the visual perception, often providing insights via
simple graphs.

The paper is organized in the following way. In the next section,
we introduce the topological concepts, i.e. a suitable wave function
topology including its invariant, which are used later on. Subsequently
we will briefly introduce the scalar paraxial wave equation, which
facilitates a highly accurate description of the dynamics of accelerated
electron waves. Finally we combine wave dynamics and topology in order
to characterize the vortex structure of the wavefield and its interaction
with atomic scattering potentials.

\section{Topological invariants of the (vortex) wave\label{sec:Topological-invariants-of}}

\begin{figure}[h]
\includegraphics[bb=100bp 280bp 480bp 560bp,clip,scale=0.5]{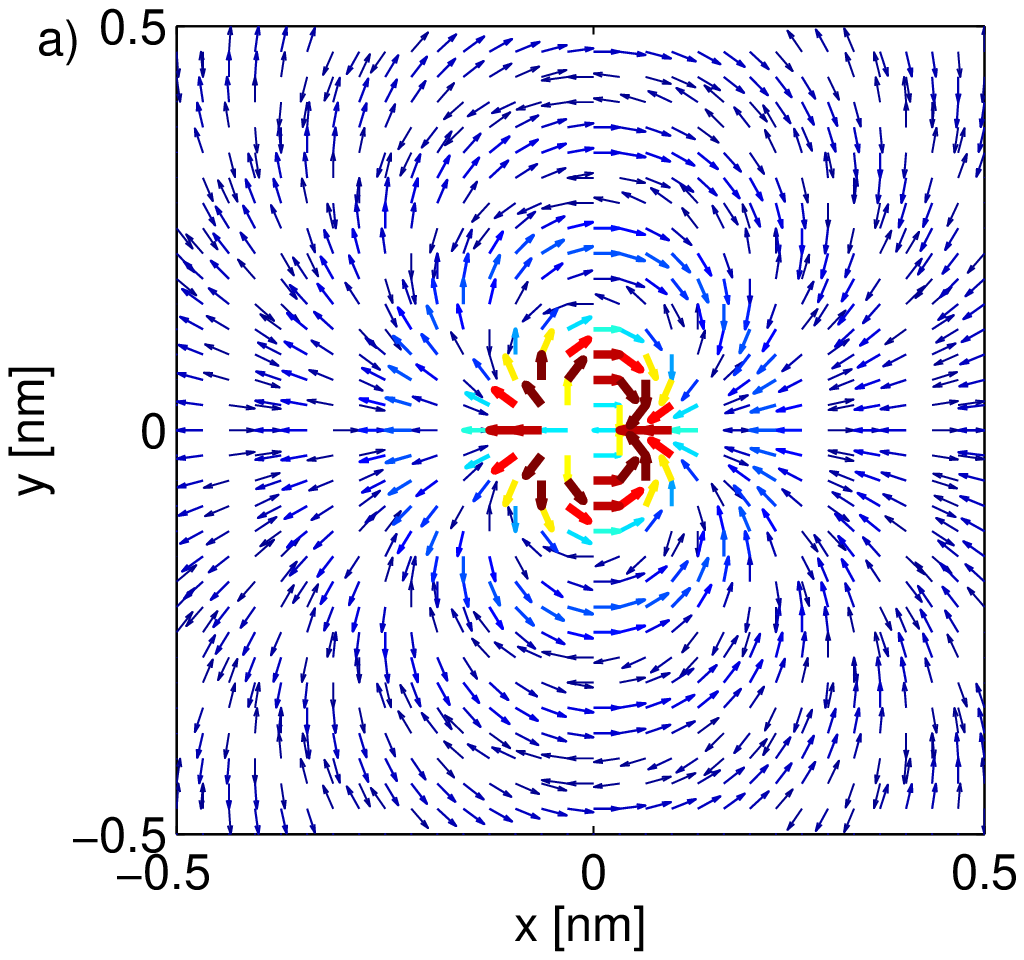}\includegraphics[bb=100bp 280bp 480bp 560bp,clip,scale=0.5]{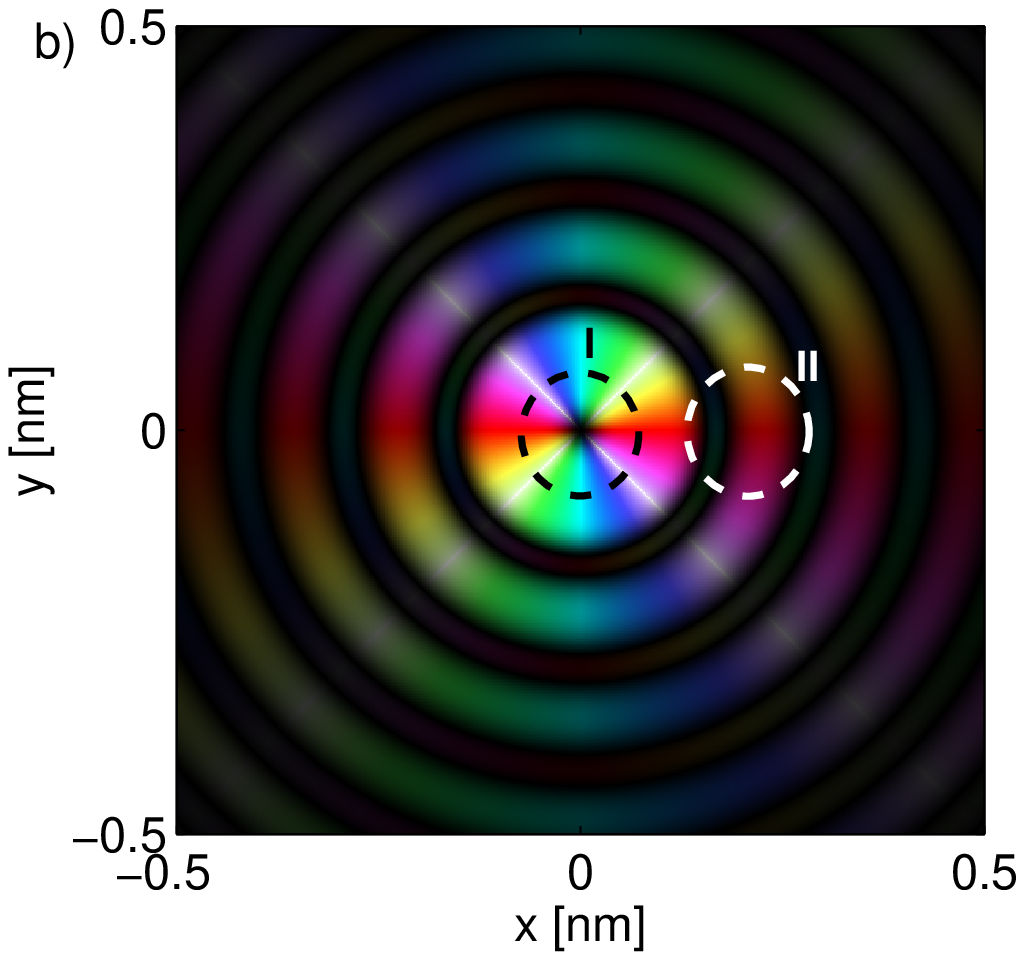}

\caption{(Color online) a) Polya map of vortex beam $\Psi_{m=2}$ (see Eq.
\ref{eq:m-beam} for details). Amplitude of the complex number is
colour-coded (light blue and thin line$\cong$small, dark red and
thick line$\cong$large). The Kronecker (Hopf) index of the phase
singularity equals $m$ and corresponds to the winding number $w$
of any closed path around the singularity. b) amplitude (value) and
phase (hue) of the same $\Psi_{m=2}$ encoded as a HSV image (domain
colouring, in the print version green$\cong$light gray, blue$\cong$dark
gray with red in between). The fully saturated white lines are used
to display constant phase lines which end at the phase singularity
by definition. \label{fig:vortex topology}}
\end{figure}

Topological invariants are features that do not change upon continuous
transformation of the wave field. Since we restrict our considerations
to the paraxial regime, that transformation \emph{is} the propagation
along the optical axis $z$ and we shall focus on the topology and
its invariants of 2D wave functions defined on planes with Cartesian
coordinates $\vec{r}_{\bot}=\left(x,y\right)^{T}$ (polar: $r,\theta$)
perpendicular to $z$. Generally, the 2D wave function is an assignment
$\Psi:\,\mathbb{R}^{2}\rightarrow\mathbb{C},\,\left(x,y\right)\rightarrow\Psi\left(x,y\right)$.
Such a map can be considered as a vector field $\left(\mathbb{\Re}\left\{ \Psi\right\} ,\Im\left\{ \Psi\right\} \right)^{T}$
composed from real and imaginary components of the complex wave. The
field might then contain 0D vortex (or phase) singularities where
the phase vorticity of the phase vector field diverges (e.g. Fig.
\ref{fig:vortex topology}a), see App. \ref{sec:Winding-number,-vorticity,}
for details). In this work we will use domain colouring (e.g. \citep{Poelke(2009)}),
i.e. mapping of the wave function $\Psi=\left|\Psi\right|\exp\left(i\varphi\right)$
into HSV (Hue, Saturation, Value) colour space (amplitude$\left|\Psi\right|\rightarrow\mathrm{V}$,
phase $\varphi\rightarrow\mathrm{H}$), to solve the non-trivial visualization
problem for the wave function (Fig. \ref{fig:vortex topology}b, in
the grayscale print version green appears as light gray, blue as dark
gray with red in between). In such maps, phase singularities show
up as prominent colour wheels elegantly revealing the $2\,\pi$ phase
jumps inherent to vortices. 

The phase singularities visible in Fig.\ref{fig:vortex topology}
are characterized by non-vanishing integer winding numbers $w\,\epsilon\,\mathbb{Z}$
of the phase gradient integrated along certain paths $2\pi w=\oint\mathrm{d}\vec{s}\cdot\vec{\nabla}\varphi$
($\mathrm{d}\vec{s}$...path element). E.g. $w$ either assumes $m$
(path I) or $0$ (path II) in Fig. \ref{fig:vortex topology}b and
paths yielding the same $w$ are equivalent under homotopy (i.e. can
be continuously transformed into each other, e.g. \citep{Frankel(1999)}).
By applying Stokes theorem 
\begin{equation}
2\pi w=\iint_{S}\mathrm{d}^{2}r_{\bot}\,\left(\vec{\nabla}\times\vec{\nabla}\varphi\right)\cdot\vec{n}=\iint_{S}\mathrm{d}^{2}r_{\bot}\,\vec{\Theta}_{p}\cdot\vec{n}\,,\label{eq:vort}
\end{equation}
where $S$ is the enclosed area with unit normal $\vec{n}$, one can
define the phase vorticity distribution $\vec{\Theta}_{p}$ which
is zero everywhere except at $0$D phase singularities. We will use
this definition to numerically locate vortices in the wave field later
on. We now define the total winding number $W$ as the phase vorticity
integrated over the whole plane, which can also be computed from the
sum of path integrals $j$ containing exactly one of the $N$ vortices,
i.e. 
\begin{equation}
W=\iint\mathrm{d}^{2}r_{\bot}\vec{\Theta}_{p}\cdot\vec{n}=\sum_{j=1}^{N}w_{j}
\end{equation}
 (e.g. \citep{Frankel(1999)}). $W$ is a topological invariant constant
under propagation because it is a continuous assignment restricted
to integers values (for an alternative derivation of the conservation
of $W$ see below, elaborate proofs can be found in the literature\citep{Frankel(1999)}).
We furthermore note two related facts about the wave function in the
vicinity of a vortex core at $\left(x_{j},y_{j}\right)^{T}$ (with
local coordinates $r_{j}=\sqrt{\left(x-x_{j}\right)^{2}+\left(y-y_{j}\right)^{2}}$
and $\theta_{j}=\arctan2\left(y-y_{j},x-x_{j}\right)$): (1) $\Psi\sim\exp\left(im\theta_{j}\right)$
which coincides with the azimuthal behaviour of a spherical harmonic
with angular momentum pointing along $z$: $Y_{l\pm l}\sim e^{\pm il\theta_{j}}$.
That leads to object entangled inelastic excitation of $\Psi\sim\exp\left(i\left(m\pm0,1\right)\theta_{j}\right)$,
which provides a suitable basis for measuring EMCD with OAM beams\citep{Verbeeck(2010),Lloyd(2012)}.
(2) The angular momentum 
\begin{equation}
-i\hbar\int\mathrm{d}^{2}r_{\bot}\,\Psi^{*}\partial_{\theta_{j}}\Psi=\hbar\int\mathrm{d}r_{j}\, r_{j}\int_{0}^{2\pi}\mathrm{d}\theta_{j}\left|\Psi\right|^{2}\partial_{\theta_{j}}\varphi=\hbar\int\mathrm{d}r_{j}\, r_{j}\oint\mathrm{d}\vec{s}\cdot\vec{\nabla}\varphi\left|\Psi\right|^{2}\label{eq:Lden}
\end{equation}
around a phase singularity has the same sign as the winding number
if the gradient of the phase along the circular path does not change
sign. That facilitates a prediction of the direction of the angular
momentum close to the vortex.

We will now briefly review the behaviour of vortices upon small perturbations
(without vortex signature) since this will reveal a classification
of possible continuous deformations of the vortex beam during propagation
along $z$ (similar considerations can be found in Ref. \citep{Berry(2007)}).
For that purpose it is sufficient to consider the lowest order Taylor
expansion of the wave around the phase singularity given by $\Psi\left(\vec{r}_{\bot},z=0\right)=\left(x+iy\right)^{m}\sim\exp\left(im\theta\right)$
(generally, an $\mathcal{O}\,\left(m\right)$ expansion is required
to describe an $m$-order vortex). We begin with the most simple perturbation
consisting of adding a linear perturbation function ($ax+bz+c$ with
$a,b,c\,\epsilon\,\mathbb{R}$ without loss of generality) to a $m=1$
vortex. In the limit of small $a$, $b$ and $c$ (continuous transformation)
\begin{equation}
\Psi'\left(\vec{r}\right)=\left(x+iy\right)+ax+bz+c+\mathcal{O}\left(\vec{r}^{2}\right)\,.
\end{equation}
Accordingly, the singularity $\Psi'\left(x_{0},y_{0}\right)=0$ gets
shifted to 
\begin{equation}
x_{0}=-\left(bz+c\right)/\left(1+a\right)\,.\label{eq:shift}
\end{equation}
A first order Taylor expansion around $x_{0}$, $y_{0}=0,z_{0}=0$
yields 
\begin{equation}
\Psi'\left(\vec{r}\right)=\left(1+a\right)x'+iy'+bz'=x''+iy'
\end{equation}
where a rotation around the $y$-axis with an angle $\alpha=\arctan\nicefrac{b}{1+a}$
was performed in the last step. Consequently, a small perturbation
added to a $m=1$ beam can only continuously shift and tilt the vortex
($\cong$ the vortex is topologically protected and the loci of phase
singularities in 3D form a line). It is interesting to note that there
is no restriction in the rotation angle $\alpha$, e.g. a 90$\text{\textdegree}$
tilt of the vortex to the side, i.e. $\Psi\left(\vec{r}\right)=x+iz+\mathcal{O}\left(\vec{r}^{2}\right)$,
is possible through continuous perturbations. The situation becomes
slightly more complex at higher-orders of $m$ due to the following
effect: Adding a small constant to the vortex, i.e. $\left(x+iy\right)^{m}+c$,
will lift the degeneracy of the zero, leading to $m$ complex zeros,
i.e. the $m>1$ beam will be transformed into $m$ first order beams,
which will then follow the perturbation rules developed above. Since
the requirement for splitting of a higher-order vortex beams is weak
(small $c$) we do not observe stable higher-order beams in the experiment
unless symmetry of the scattering potential prevents the $m$-fold
splitting. In the light of the last discussion, a higher-order vortex
can be regarded as a degenerate superposition of nondegenerate first
order vortices where the details of degeneracy lifting can be analyzed
by topological arguments\cite{Freund(1999)}. Similarly, also non-collinear
crossing of vortex lines, e.g. $\Psi\left(\vec{r}\right)=\left(x+iz\right)\left(y+iz\right)+\mathcal{O}\left(\vec{r}^{3}\right)$,
can occur in the wave field. In total, we can identify 3 main effects
upon perturbation: shift, tilt and (non-)collinear crossing. All of
which can occur upon paraxial propagation of the wave and we will
show in Sec. \ref{sec:Paraxial-wave-dynamics} that the simple perturbation
scheme discussed above is indeed sufficient to describe the possible
modulation of vortices. Note that none of the above effects destroys
a vortex line similar to the 2nd Helmholtz theorem in fluid mechanics\cite{Lighthill(1986)}.
Consequently also the total number of vortices $W$ remains constant
corroborating the proof sketch above. By evaluating the location of
the vortex loci upon propagation we can detect the vortex line skeleton,
hence the topological structure of the wave function.

\section{Paraxial wave dynamics\label{sec:Paraxial-wave-dynamics}}

The scalar paraxial wave equation governing the propagation of a fast
electron wave inside a transmission electron microscope (TEM) reads\citep{Graef(2003)a}%
\footnote{Shortcomings of the paraxial approximation as well as vector potentials
will be considered elsewhere.%
} 
\begin{equation}
\partial_{z}\Phi\left(\vec{r}\right)=i\left(\frac{EV\left(\vec{r}\right)}{k_{0}\hbar^{2}c^{2}}+\frac{1}{2k_{0}}\Delta_{\bot}\right)\Phi\left(\vec{r}\right)\,,\label{eq:parax}
\end{equation}
thus can be considered as a time dependent 2D Schrödinger equation
($k_{0}...$magnitude of wave vector, $E$...energy of electron wave,
$V$...electrostatic scattering potential). Consequently, we can analyze
commutator relations of certain operators with the Hamilton operator,
determining whether they are conserved or not upon propagation. In
case of OAM the commutator $\left[L_{z},H\right]=\left[-i\hbar\partial_{\theta},V\right]$
only vanishes if $V$ does not depend on $\theta$, i.e. OAM is not
conserved upon propagation (contrary to the total winding number)
and can even change sign (see \citep{Loffler(2012)} for explicit
numerical calculation). This can be also understood by noting the
paraxial equation in the rotating basis $\Phi\left(\vec{r}\right)=\chi\left(\vec{r}\right)\exp\left(im\theta\right)$
\begin{eqnarray}
\partial_{z}\chi\left(\vec{r}\right) & = & i\left(\frac{EV\left(\vec{r}\right)}{k_{0}\hbar^{2}c^{2}}+\frac{1}{2k_{0}}\Delta_{\bot}-\frac{m^{2}}{2k_{0z}r^{2}}+\frac{im}{k_{0}\hbar r^{2}}\hat{L}_{z}\right)\chi\left(\vec{r}\right)\,,\label{eq:radial_parax}
\end{eqnarray}
where a non-vanishing last term leads to a destruction of the initial
OAM eigenstate. The propagation of an electron wave along $z$ can
be approximated by successive free space Fresnel propagation $\Phi\left(\vec{r}_{\bot},z+\delta z\right)=\left(1+i\delta z/\left(2k_{0}\right)\Delta_{\bot}\right)\Phi\left(\vec{r}_{\bot},z\right)$
and projected potential phase shifts $\Phi\left(\vec{r}_{\bot},z+\delta z\right)=\left(1+iC_{E}V_{p}\left(\vec{r}_{\bot}\right)\right)\Phi\left(\vec{r}_{\bot},z\right)$
($C_{E}=\frac{E}{k_{0}\hbar^{2}c^{2}}$, $V_{p}\left(\vec{r}_{\bot}\right)=\int_{z}^{z+\delta z}\mathrm{d}z\, V\left(\vec{r}\right)$)
for small propagation steps $\delta z$. The corresponding numeric
integration scheme is referred to as Multislice\citep{Cowley(1957)})
and will be used below. 

We will now show that the previously identified generic situations
(i.e. shift, bending, splitting) are locally compatible with the paraxial
equation, which will effectively impose certain constraints on the
perturbation parameters $a$, $b$ and $c$. Since solutions with
vortices oriented along $z$ are well known it remains to discuss
the horizontal orientation of vortex lines. Indeed, the horizontal
vortex line with the local shape 
\begin{equation}
\Psi\left(\vec{r}_{\bot},z=0\right)=x+2k_{0}\left(ax^{2}+by^{2}\right)+\mathcal{O}\left(r^{3}\right)\,,
\end{equation}
and second order term coefficients $a+b=1$ lead to the required $iz$
dependency when propagated at small distances. The superposition of
perpendicular (non-collinear) vortex lines, e.g. the combination of
2 vortex lines oriented along $y$ and $x$ to $\Psi\left(\vec{r}\right)=\left(x+iz\right)\left(y+iz\right)+\mathcal{O}\left(\vec{r}^{3}\right)$
is also compatible with the paraxial equation if the wave function
\begin{equation}
\Psi\left(\vec{r}_{\bot},z=0\right)=xy+2k_{0}\left(ax^{3}+bx^{2}y+cy^{2}x+dy^{3}\right)+\mathcal{O}\left(\vec{r}^{4}\right)
\end{equation}
 with $6a+2c=1$ and $6d+2b=1$. The latter result shows that non-collinear
crossing of vortex lines is not forbidden in a paraxial wave field. 

We will now analyze the motion of a vortex (see Ref. \cite{Rozas(1997)}
for similar considerations for optical vortices): In line with the
integration scheme noted above we distinguish between (A) phase modulations
such as produced by a weak atomic potential $\Phi=\left(1+iC_{E}V_{p}\right)\left(x+iy\right)$
and (B) amplitude modulations as occurring under free space propagation
($\cong$interference) $\Phi=\left(1+Ax\right)\left(x+iy\right)$.
The infinitesimal paraxial free space propagation step (see Eq. \ref{eq:parax})
after phase modulation reads 
\begin{eqnarray}
\Phi\left(\vec{r}_{\bot},z+\delta z\right) & = & \left(1+i\nicefrac{\delta z}{2k_{0}}\Delta\right)\left(1+iC_{E}V_{p}\left(\vec{r}_{\bot}\right)\right)\left(x+iy\right)\nonumber \\
 & = & \left(1+iC_{E}V_{p}\left(\vec{r}_{\bot}\right)-\nicefrac{\delta z}{2k_{0}}C_{E}\Delta_{\bot}V_{p}\left(\vec{r}_{\bot}\right)\right)\left(x+iy\right)-\nicefrac{\delta z}{k_{0}}\left(\partial_{x}+i\partial_{y}\right)V_{p}\left(\vec{r}_{\bot}\right)
\end{eqnarray}
and we recognize a $x\,\left(y\right)$-shift proportional to $\partial_{x}V_{p}\,\left(\partial_{y}V_{p}\right)$
(see Eq. (3)). Thus, if the vortex is placed in the vicinity of a
radial atomic potential it is attracted towards the positive screened
Coulomb potential of an atom. In case of an amplitude modulation one
obtains 
\begin{equation}
\Phi\left(\vec{r}_{\bot},z+\delta z\right)=\left(1+i\nicefrac{\delta z}{2k_{0}}\Delta\right)\left(1+Ax\right)\left(x+iy\right)=\left(1+Ax\right)\left(x+iy\right)+i\nicefrac{\delta z}{2k_{0}}A\,,
\end{equation}
i.e. the vortex is shifted perpendicular to the modulation direction
which will superimpose on phase modulation to yield the total motion
of the vortex line. Explicit examples of that behaviour will be shown
by numerically propagating a vortex wave through a model scattering
potential. We have deliberately picked a SrTiO$_{3}$ single crystal
because it facilitates the simulation of a variety of scattering situations
due to the combination of heavy and light atomic species in a relatively
simple perovskite lattice. Furthermore, the material resembles a large
class of perovskite oxides with e.g. interesting (multi-)ferroic properties
which might be probed by EMCD experiments utilizing vortex beams in
the near future. The crystal was oriented in $\left[001\right]$-direction
to provide well-aligned atomic columns. An incoming vortex beam of
order $m$ as produced by the commonly used fork aperture in the condenser
aperture of a TEM in scanning mode (STEM) has the general shape of
a generalized hypergeometric function 
\begin{eqnarray}
\Phi_{m}\left(r_{\bot},\theta,z=0\right) & = & e^{im\theta}\int_{0}^{\alpha}J_{m}(kr)kdk\nonumber \\
 & = & e^{im\theta}\frac{2^{-m}r_{\bot}^{m}k_{max}^{2+m}}{(2+m)\Gamma\left(1+m\right)}\,_{1}F_{2}\left(1+\frac{m}{2};2+\frac{m}{2},1+m;-\frac{1}{4}r_{\bot}^{m}k_{max}^{2}\right)\label{eq:m-beam}
\end{eqnarray}
when focused on the crystal entrance face ($k_{max}=k_{0}\alpha_{max}$...aperture
size, see App. \ref{sec:Analytic-expression-for} for details). In
order to facilitate a straight forward discussion of the characteristic
vortex propagation features, we restricted our analysis to $m=1,2,4$
vortex beams exactly centered on and slightly off-center ($0.04$
nm) particular positions, namely the TiO-column and the O-column.%
\footnote{The Sr column results resemble that of the TiO column due to the similar
scattering power. %
} Off-centering was investigated in order to consider realistic experimental
disturbance of the beam position introduced by instrument limitations,
specimen drift and thermal motion. Parameters of the simulation separated
into microscopical (A), material (B) and numerical (C) are: (A) $200$
kV acceleration voltage, $15$ mrad illumination aperture, (B) SrTiO$_{3}$(SG=$\mathrm{Pm}\overline{3}\mathrm{m}$,$a=0.395$
nm), $\left[001\right]$-orientation, $20$ nm thickness, (C) Supercell
size to suppress artifacts from periodic boundary conditions: 7x7
$a\,\cong$ ($2.765\times2.765$ nm), 560x560 sampling, $\delta z=0.1975$
nm propagation step, atomic scattering potentials from Ref. \cite{Weickenmeier(1991)}.
The domain coloured wave will be depicted at different $z$-coordinates
and vortex lines will be drawn to illustrate the motion of phase singularities.
The vortex lines have been determined by detecting singularities in
the phase vorticity (Eq. \ref{eq:vort}) and connecting adjacent detected
singularities along $z$. The numerical accuracy of the vortex detection
is limited by sampling in both the horizontal ($r_{\bot}$) and vertical
($\delta z$) domain, which becomes particularly apparent at the turning
points of the loops which are frequently not completely resolved.

\begin{figure}[h]
\includegraphics[bb=0bp 440bp 595bp 842bp,clip,scale=0.7]{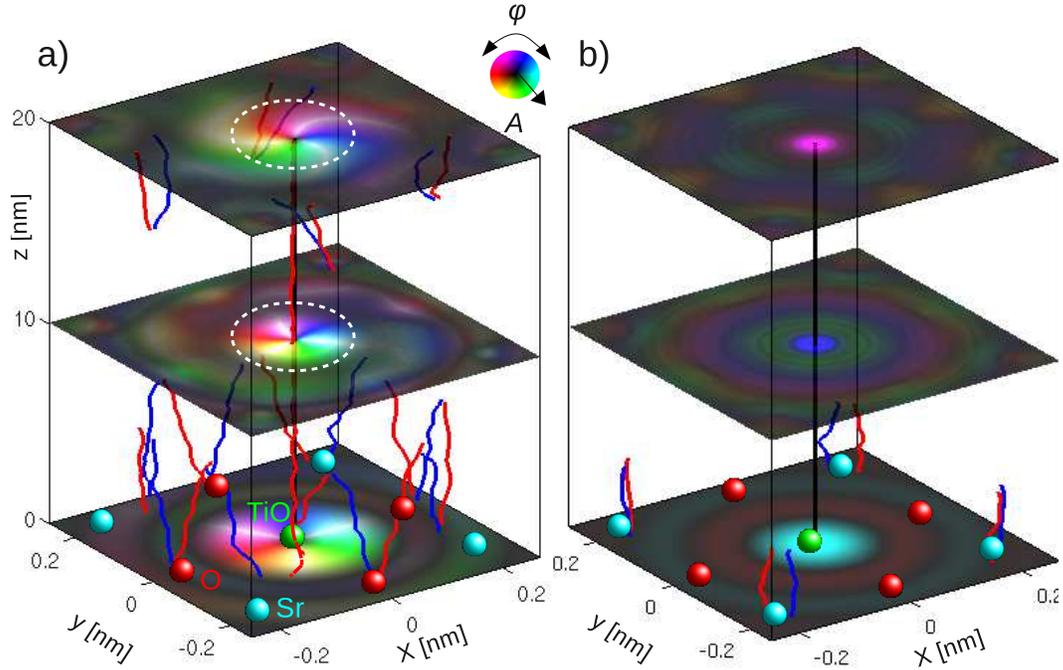}\caption{(Color online) a) $m=1$ vortex and b) $m=0$ conventional beam. The
wave function at different $z$-planes are displayed domain coloured.
$m=1\,(-1)$ vortex lines are blue/dark gray (red/light gray). Center
of mass line is black. The atomic columns of SrTiO$_{3}$ are indicated
at the bottom plane. The undisturbed vortex region is indicated by
a white circle in a). Note the closed zig-zag ring formed by red/light
gray and blue/dark gray lines at the bottom of a). \label{fig:m=00003D00003D1}}
\end{figure}

\begin{figure}[h]
\textcolor{black}{\includegraphics[scale=0.5]{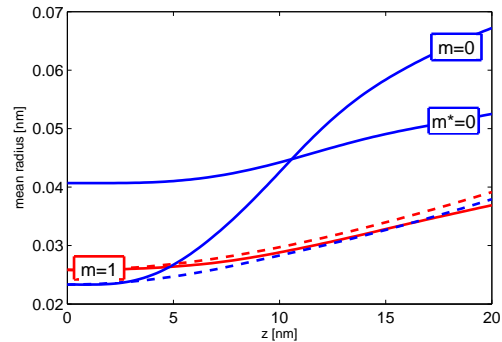}}

\textcolor{black}{\caption{(Color online) Delocalization upon propagation of STEM beams of two
$m=0$ (blue/dark gray, for comparison $m^{*}$ had a strongly reduced
opening angle (i.e. larger initial delocalization) of 5 mrad) and
one $m=1$ (red/light gray) centered on TiO column. Dashed curves
of free propagating beams (without any scattering potential) are shown
to illustrate the delocalization effect due to scattering. The $m=1$
vortex beam shows a significantly reduced delocalization compared
to normal $m=0$ beam with same convergence angle. \label{fig:Delocalization-of-STEM}}
}

\end{figure}

In Fig. \ref{fig:m=00003D00003D1}a we show the vortex line of a $m=1$
beam centered on the TiO-column. The central vortex line is straight
and one can distinguish two regions around the vortex: Within the
white circle the wave function corresponds well to an angular momentum
eigenstate $\sim\exp\left(im\theta\right)$. Outside, the equiphaselines
are strongly bent, which coarsely resembles a Rankine vortex behaviour\cite{Acheson(1990)}.
The radius of the circle is determined by strong deviations from the
rotationally symmetric scattering potential given by neighboring atomic
columns, i.e. the last term of the rotated paraxial Eq. \ref{eq:radial_parax}
remains small only within the radius. To compare the delocalization
upon scattering of vortex and non-vortex (conventional) beams we plot
the mean radius $\overline{r}\left(z\right)=\left(\int\mathrm{d}^{2}r_{\bot}\Phi^{*}\left(\vec{r}\right)r_{\bot}^{2}\Phi\left(\vec{r}\right)\right)^{1/2}$
of various beams in Fig. \ref{fig:Delocalization-of-STEM}. Here we
note that the presence of a stable topologically protected zero centered
on the strongly scattering atomic column prevents the vortex beam
from scattering (and delocalizing) rapidly. In the example shown the
mean radius even increases less slowly than the freely propagating
beam due to the attracting force of the positive atom cores in the
column. That behaviour stays in contrast to the corresponding (i.e.
with same convergence angle) conventional $m=0$ beam (Fig.\ref{fig:m=00003D00003D1}b),
which initially has his maximum at the column. At the same convergence
angle the conventional beam is initially slightly more localized than
the vortex beam but delocalizes significantly stronger upon scattering,
both compared to the vortex and the freely propagating conventional
beam. We remark that the quantitative details of the delocalization
reduction of the vortex beam depend on a large set of parameters,
like acceleration voltage, convergence angle, scattering potential,
specimen thickness, hence have to be worked out for each particular
problem in practice. 

\textcolor{black}{We} furthermore observe a formation of numerous
vertical vortex loops upon propagation, visible as vortex-antivortex
pairs in a particular $z-$plane. They are predominantly attached
to atomic columns and have been known for a long time to occur in
conventional beams scattered at atomic potentials (Fig. \ref{fig:m=00003D00003D1}b)\citep{Allen(2001)c}.
Their presence visually explains the local sign flips in the local
angular momentum density (see Eq. \ref{eq:Lden}), observed in Ref.
\citep{Loffler(2012)}. Less frequently we also observed predominantly
horizontally oriented loops. E.g. the vortex-antivortex pairs emerging
from the oxygen column in Fig. \ref{fig:m=00003D00003D1}a form a
horizontal zig-zag loop enclosing the central vortex. However, we
could not detect any knot structure (i.e. linked loops) in the wave
field as reported for photonic wave fields\citep{Leach(2004)}. Thus
they seem to be a less probable structure although they are not forbidden
in general.

\begin{figure}[h]
\includegraphics[bb=0bp 440bp 595bp 842bp,clip,scale=0.7]{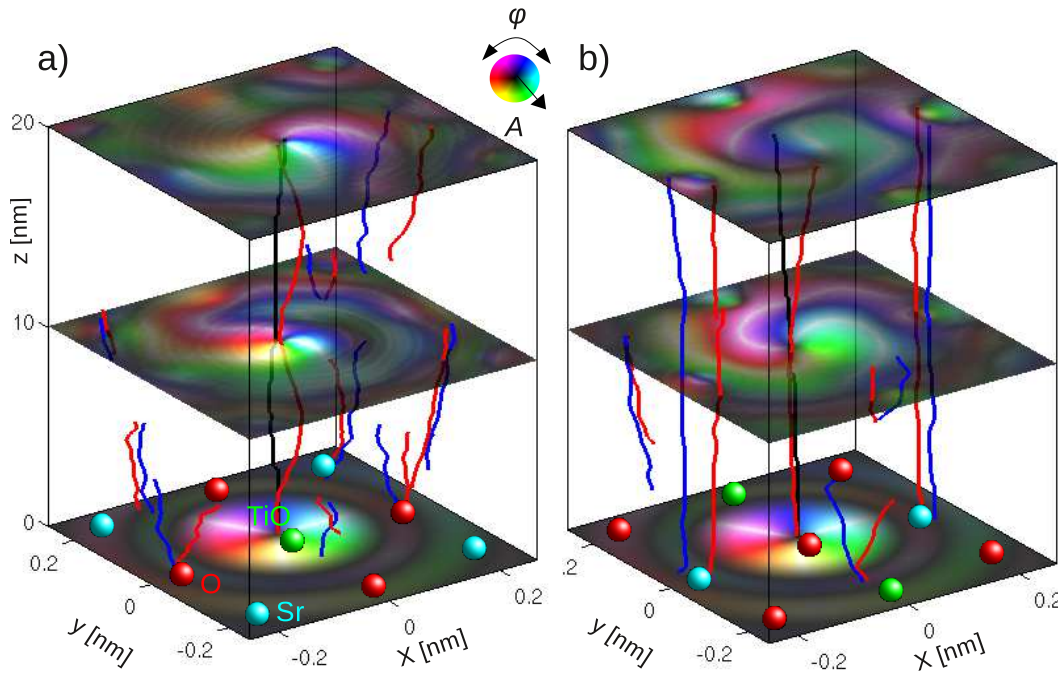}\caption{(Color online) $m$=1 vortex beam slightly off-center a) the TiO and
b) the O column. The wave function at different $z$-planes are displayed
domain coloured. $m=1\,(-1)$ vortex lines are blue/dark gray (red/light
gray). Center of mass line is black. \label{fig:m=00003D00003D0}}
\end{figure}

Besides vortex loops the off-center $m=1$ beam (Fig. \ref{fig:m=00003D00003D0}a)
also shows circulation of the central singularity around the TiO column.
That circular motion is the sum of a phase modulation attracting the
vortex towards the atomic column and an amplitude modulation (focusing
property of positive atomic potential) forcing an angular shift (see
above). It is not observed when tracking the center of mass (see Fig.
\ref{fig:m=00003D00003D0}a). In agreement with Eq. \ref{eq:shift}
the spiral direction changes when the topological charge changes sign.
If the beam is centered on the O column the circulation period is
increased as a consequence of the smaller scattering potential of
the atomic column (Fig. \ref{fig:m=00003D00003D0}b). 

\begin{figure}[h]
\includegraphics[bb=0bp 440bp 595bp 842bp,clip,scale=0.7]{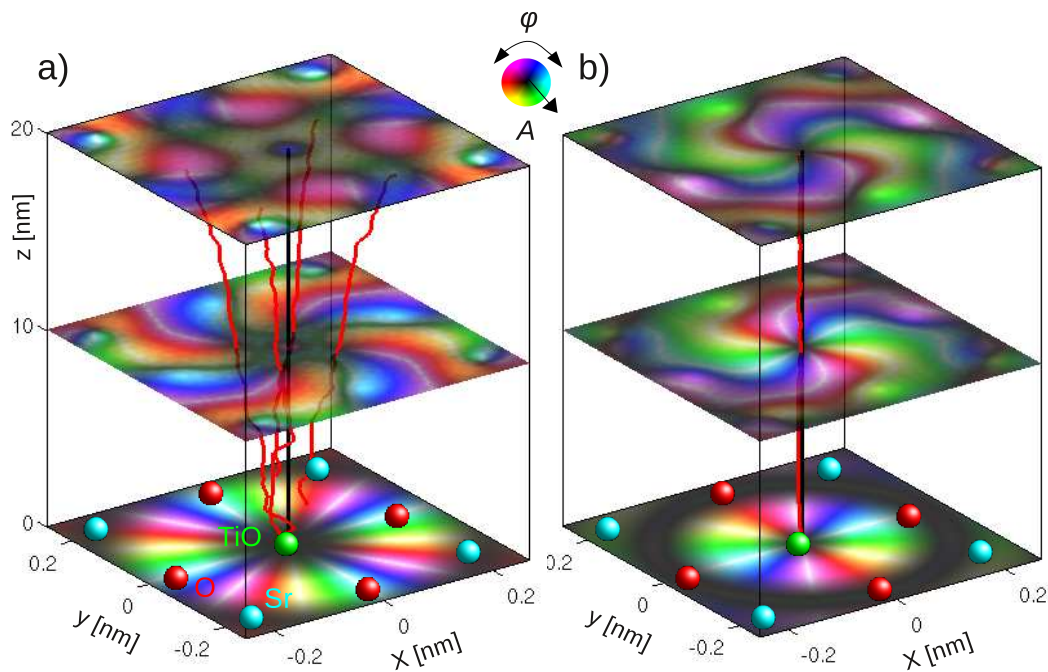}\caption{(Color online) a) $m=4$ and b) $m=2$ vortex beam centered at TiO
column. The wave function at different $z$-planes are displayed domain
coloured. $m=1\,(-1)$ vortex lines are blue/dark gray (red/light
gray). Center of mass line is black.\label{fig:m=00003D00003D2}}
\end{figure}

In case of higher order beams $m>1$ we additionally expect to observe
splitting into $m$ first order beams (see above). We found a restriction
to this behaviour depending of the site symmetry around the beam:
At positions without rotational symmetry the higher order beams split
into $m$ first order beams. At positions with local $n$-fold rotational
symmetry the beam split into $k\epsilon\mathbb{N}\cdot n$ first-order
$m=\mathrm{sign}\left(m\right)1$ vortices arranged symmetrically
around one $\mathrm{sign}\left(m\right)\left(\left|m\right|-k\cdot n\right)$
vortex with $k$ equaling the down-rounded $m/n$. Consequently, if
$m<n$, the $m$-order beam is protected from splitting. E.g. a $m=4$
beam starts to split into 4 off-center $m=1$ when centered at the
4fold symmetric TiO column (Fig. \ref{fig:m=00003D00003D2}a), whereas
the $m=2$ beam stays stable (Fig. \ref{fig:m=00003D00003D2}b). Similarly
a $m=6$ beam splits into $4$ off-center $m=1$ and one centered
$m=2$ beam. We emphasize, however, that perfect $n$-fold rotational
site symmetry is never fulfilled in practice, due to e.g. a non-perfect
centering of the beam, imperfect apertures or the thermal motion of
the atoms. In practice, these perturbations lead to the splitting
of \emph{all} high-order beams into first order beams and we observe
(up to $m=6$) the following modified splitting for a $m$-vortex
on a approximately $n$-fold symmetric site into $k\epsilon\mathbb{N}\cdot n$
symmetric off-center $m=\mathrm{sign}\left(m\right)1$vortices and
$\left|m-k\cdot n\right|$center $m=\mathrm{sign}\left(m-k\cdot n\right)1$
vortices with: 
\begin{equation}
k=\begin{cases}
\begin{array}{c}
\left\lfloor \left|m\right|/n\right\rfloor \\
\left\lceil \left|m\right|/n\right\rceil 
\end{array} & \begin{array}{c}
\mathrm{if}\,\mathrm{mod}\left(\left|m\right|,n\right)\leq n/2\\
\mathrm{if}\,\mathrm{mod}\left(\left|m\right|,n\right)\geq n/2
\end{array}\end{cases}
\end{equation}
Here, $\left\lfloor .\right\rfloor $ ($\left\lceil .\right\rceil $)
denote rounding down (up). E.g. the $m=3$ centered on the 4-fold
symmetric TiO column, protected under perfect symmetry, now splits
into 4 off-center $m=1$ and 1 centered $m=-1$ vortex. Similarly,
additional vortex-antivortex pairs with a symmetry adapted distribution
may occur in the wave field (not shown in Fig. \ref{fig:m=00003D00003D2}).

\section{Summary, Discussion}

We have analyzed the fabric of vortex lines occurring in a vortex
beam upon propagation through an atomic lattice with respect to general
topological information. The following features with corresponding
significance could be distinguished: (I) A Rankine like vortex structure
with significantly reduced diffusion of the vortex beam compared to
a non-vortex beam, when centered on an atomic column. Therefore vortex
beams hold the promise to significantly improve atomic resolution
STEM-EELS techniques, where the ubiquitous delocalization of the probe
upon propagation systematically obscures the atomic information\textcolor{black}{\citep{Egerton(1996)}.}
(II) A systematic deflection of vortex lines from amplitude and phase
perturbations leading e.g. to a circulation of vortex lines around
atomic columns depending on the sign of the vorticity. (III) Symmetry
dependent splitting of $m$-order vortex beams. We showed that combining
topological concepts (conservation of total winding number) and symmetry
allows predicting the vortex line structure of the beam without solving
the scattering equations at all. In particular one can predict which
high-order vortex beam will maintain a centered topologically protected
zero under certain rotational symmetries. (IV) The appearance of vortex-antivortex
pairs corresponding to not-knotted vortex loops predominantly attached
to atomic columns. That observation is important for the design of
phase unwrapping algorithms for electron holography, where the linking
of phase singularities represents the major obstacle\citep{Ghiglia(1998)}.
We furthermore point out that the 3D vortex line behaviour is useful
to classify vortex beam creators in general. The closeness of vortex
lines implies that in order to produce a vortex one has to create
a vortex-antivortex pair in the first place and before singling out
later on one of these. This can be done in a symmetric fashion, e.g.
fork\citep{Verbeeck(2010)} and spiral\citep{Verbeeck(2012)} aperture
or in an unsymmetrical fashion, e.g. the finite thickness phase plate\citep{Uchida(2010)}.

\appendix

\section{Winding number, vorticity, angular momentum\label{sec:Winding-number,-vorticity,}}

In the following we note some definitions partly utilized in the text
and frequently used in the context of vortex physics. The winding
number can be obtained from various definitions 
\begin{equation}
2\pi w\equiv\oint\vec{\nabla}\varphi\cdot\mathrm{d}\vec{s}=-i\oint\frac{\vec{\nabla}\Psi}{\Psi}\cdot\mathrm{d}\vec{s}=\iint_{S}\vec{\nabla}\times\vec{\nabla}\varphi\cdot\vec{n}\mathrm{d}^{2}r\,,
\end{equation}
where the kernel of the last surface integral defines the phase vorticity
distribution 
\begin{equation}
\vec{\Theta}_{p}\equiv\vec{\nabla}\times\vec{\nabla}\varphi
\end{equation}
used in the text. The vorticity is defined in the following way 
\begin{eqnarray}
\vec{\Theta} & \equiv & \vec{\nabla}\times\vec{v}=\frac{1}{m}\vec{\nabla}\times\vec{j}\\
 & = & \frac{-i\hbar}{2m^{2}}\vec{\nabla}\times\left(\Psi^{*}\vec{\nabla}\Psi-\Psi\vec{\nabla}\Psi^{*}\right)\nonumber \\
 & = & \frac{-i\hbar}{m^{2}}\vec{\nabla}\Psi^{*}\times\vec{\nabla}\Psi=\frac{2\hbar}{m^{2}}\vec{\nabla}\Re\times\vec{\nabla}\Im\nonumber \\
 & = & \frac{-i\hbar}{m^{2}}\vec{\nabla}\times\left(\rho\vec{\nabla}\varphi\right)\nonumber \\
 & = & \frac{-i\hbar}{m^{2}}\left(\vec{\nabla}\rho\times\vec{\nabla}\varphi+\rho\vec{\nabla}\times\vec{\nabla}\varphi\right)\,.
\end{eqnarray}
The last line demonstrated that vorticity should not be confused with
the phase vorticity distribution. The angular momentum density reads

\begin{equation}
\vec{L}=m\vec{r}\times\vec{j}=\frac{-i\hbar}{2}\vec{r}\times\left(\Psi^{*}\vec{\nabla}\Psi-\Psi\vec{\nabla}\Psi^{*}\right)
\end{equation}
and is only directly proportional to the vorticity in case of an angular
momentum eigenstate.

\section{Paraxial equation in a rotating basis\label{sec:Paraxial-equation-in}}

The paraxial wave equation in a rotated basis used in the text (Eq.
\ref{eq:radial_parax}) is found by performing the following substitution
in the standard paraxial equation: $\Phi\left(\vec{r}\right)=\chi\left(\vec{r}\right)\exp\left(im\theta\right)$
\begin{eqnarray}
\partial_{z}\Phi\left(\vec{r}\right) & = & i\left(\frac{EV\left(\vec{r}\right)}{k_{0z}\hbar^{2}c^{2}}+\frac{1}{2k_{0z}}\Delta_{\bot}\right)\Phi\left(\vec{r}\right)\\
 & = & i\left(\frac{EV\left(\vec{r}\right)}{k_{0z}\hbar^{2}c^{2}}\Phi\left(\vec{r}\right)+\frac{\exp\left(im\theta\right)}{2k_{0z}}\left(\Delta_{\bot}\chi\left(\vec{r}\right)-\frac{m^{2}}{r_{\bot}^{2}}+\frac{2im}{r_{\bot}^{2}}\left(x\partial_{y}-y\partial_{x}\right)\chi\left(\vec{r}\right)\right)\right)\,,\nonumber 
\end{eqnarray}
which leads to the following differential equation for the non-rotating
$\chi$: 
\begin{eqnarray}
\partial_{z}\chi\left(\vec{r}\right) & = & i\left(\frac{EV\left(\vec{r}\right)}{k_{0z}\hbar^{2}c^{2}}+\frac{1}{2k_{0z}}\Delta_{\bot}-\frac{m^{2}}{2k_{0z}r_{\bot}^{2}}+\frac{im}{k_{0z}\hbar r_{\bot}^{2}}\hat{L}_{z}\right)\chi\left(\vec{r}\right)\,.
\end{eqnarray}
The last term in the bracket is missing for OAM eigenstates because
there $\chi\left(\vec{r}\right)=\chi\left(r_{\bot},z\right)$.

\section{Analytic expression for paraxial vortex beams\label{sec:Analytic-expression-for}}

To calculate the initial shape of the vortex beam at the sample plane
formed in the microscope equipped with a fork aperture we need to
consider first the aperture plane (radius $Q$) and subsequently propagate
the solution to the specimen plane. In the fork aperture plane$\Psi=\Theta\left(q-Q\right)\left|\exp\left(i\frac{k_{0}}{2}q\right)+\exp\left(-i\frac{k_{0}}{2}q\right)\exp\left(im\varphi\right)\right|^{2}=\Theta\left(q-Q\right)\left(2+\exp\left(ik_{0}q\right)\exp\left(-im\varphi\right)+\mathrm{c.c.}\right)$
assuming a non binary fork aperture (binarization mainly introduces
higher-order vortices, which we neglect here). The wave function in
the specimen plane, obtained by Fourier transformation, consists of
an Airy shaped center band ($m=0$), and 2 complex conjugate sideband
($\pm m$) with 
\begin{eqnarray}
\Psi_{m}\left(\vec{r}_{\bot}\right) & = & \int d^{2}q\Theta\left(q-Q\right)\exp\left(im\varphi\right)\exp\left(i\vec{q}\vec{r}_{\bot}\right)\\
 & = & \int_{0}^{Q}\int_{0}^{2\pi}dqd\varphi q\exp\left(im\varphi\right)\exp\left(iqr\cos\left(\theta-\varphi\right)\right)\nonumber \\
 & = & 2\pi i^{m}\exp\left(im\theta\right)\int_{0}^{Q}dqqJ_{m}\left(qr_{\bot}\right)\nonumber \\
 & = & \exp\left(im\theta\right)\frac{2\pi i^{m}2^{-m}r_{\bot}^{m}Q^{2+m}}{(2+m)\Gamma\left(1+m\right)}\,_{1}F_{2}\left(1+\frac{m}{2};2+\frac{m}{2},1+m;-\frac{1}{4}r_{\bot}^{2}Q^{2}\right)\,,\nonumber 
\end{eqnarray}
where $_{p}F_{q}(a;b;z)$ is the generalized hypergeometric function\citep{Abramowitz(1964)}.
Accordingly, a vortex beam with the characteristic phase $\exp\left(im\theta\right)$
is obtained by cutting out one sideband. It has to be noted, however,
that the support of $_{1}F_{2}(a;b;z)$ is not limited, i.e. the sidebands
reach into the center band and vice versa. That effect and the binary
character of typical fork apertures leads to additional modifications
to $\Psi_{m}$, which will be neglected here. For the relevant cases
of $m=1,2,4$ used in this work the expression simplifies to 
\begin{equation}
\Phi_{m=1}\left(\vec{r}_{\bot}\right)=\frac{i\pi^{2}k_{max}}{r_{\bot}}\exp\left(im\theta\right)\left(J_{1}\left(k_{max}r_{\bot}\right)H_{0}\left(k_{max}r_{\bot}\right)-J_{0}\left(k_{max}r_{\bot}\right)H_{1}\left(k_{max}r_{\bot}\right)\right)\,,
\end{equation}
\begin{equation}
\Phi_{m=2}\left(\vec{r}_{\bot}\right)=\frac{-2\pi}{r_{\bot}^{2}}\exp\left(im\theta\right)\left(2-2J_{0}\left(k_{max}r_{\bot}\right)-Qr_{\bot}J_{1}\left(k_{max}r_{\bot}\right)\right)
\end{equation}
and 
\begin{equation}
\Phi_{m=4}\left(\vec{r}_{\bot}\right)=\exp\left(im\theta\right)\frac{4r_{\bot}+\left(r_{\bot}^{2}k_{max}-\frac{8}{k_{max}}\right)J_{1}\left(k_{max}r_{\bot}\right)-8r_{\bot}J_{2}\left(k_{max}r_{\bot}\right)}{r_{\bot}^{3}}\,,
\end{equation}
where $J$ denotes Bessel functions of the first kind and $H$ denotes
Struve functions.

\end{document}